\documentclass[modern]{aastex61}

\usepackage{graphicx}
\usepackage{amsmath}
\usepackage{amssymb}
\usepackage{enumitem}

\newcommand\mybar{\kern1pt\rule[-\dp\strutbox]{.8pt}{\baselineskip}\kern1pt}

\setlist[itemize]{noitemsep, topsep=0pt, leftmargin=*}

\shorttitle{``Down to Earth'' Limits on Unidentified Aerial Phenomena}
\shortauthors{Loeb}

\begin{document}

\title{``Down to Earth'' Limits on Unidentified Aerial Phenomena}

\author{Abraham Loeb}
\affiliation{Astronomy Department, Harvard University, 60 Garden
  St., Cambridge, MA 02138, USA}

\begin{abstract}
A recent report by astronomers about Unidentified Aerial Phenomena
(UAP) in Ukraine (arXiv:2208.11215) suggests dark phantom objects of size 3--12 meters,
moving at speeds of up to 15 km~s$^{-1}$ at a distance of up to
10--12~km with no optical emission.  I show that the friction of such
objects with the surrounding air would have generated a bright optical
fireball. Reducing their inferred distance by a factor of ten is fully
consistent with the size and speed of artillery shells.
\end{abstract}

\section{Introduction}

Recently, astronomers in Ukraine reported about Unidentified Aerial
Phenomena (UAP) that fall into two categories: bright and
dark~\citep{Zhil22}. The dark objects with no optical emission were
labeled as ``Phantoms''. They were characterized by a size of 3--12
meters and speeds up to 15~km~s$^{-1}$ at a distance of up to
10--12~km. If real, such objects exceed the capabilities of human-made
aircrafts or rockets. Here, I show that the distance of these dark
objects must have been incorrectly overestimated by an order of
magnitude, or else their bow shock in the Earth's atmosphere would
have generated a bright fireball with an easily detectable optical
luminosity.

The interest in UAP stems from their potential non-human
origin. Extraterrestrial equipment could arrive in two forms: space
trash, similar to the way our own interstellar probes (Voyager 1 \& 2,
Pioneer 10 \& 11 and New Horizons) will appear in a billion years, or
functional equipment, such as autonomous devices equipped with
Artificial Intelligence (AI). The latter would be an ideal choice for
crossing the tens of thousands of light years that span the scale of
the Milky Way galaxy and could survive even if the senders are not
able to communicate.

It is likely that any functional devices embedded in the Earth's
atmosphere are not carrying biological entities because these would
not survive the long journey through interstellar space and its harsh
conditions, including bombardment by energetic cosmic-rays, X-rays and
gamma-rays~\citep{Ho1,Ho2,Ho3}. Interstellar gas and dust particles
deposit a kinetic energy per unit mass that exceeds the output of
chemical explosives at the speed of tens of km~s$^{-1}$ characterizing
rockets.  However, technological gadgets with AI can be shielded to
withstand the hazards of space, repair themselves mechanically, or
even reproduce given the resources of a habitable planet like
Earth. With Machine Learning capabilities, they can adapt to new
circumstances and pursue the goals of their senders without any need
for external guidance.

As argued by John von Neumann in 1939, the number of such devices
could increase exponentially with time if they
self-replicate~\citep{Freitas80}, a quality enabled by 3D printing and
AI technologies. Physical artifacts might also carry messages, as
envisioned by Ronald Bracewell in 1960~\citep{Bracewell60,Valdes85}.

\section{Propulsion Methods}

In principle, the fastest gadgets could be launched by lightsails,
pushed by powerful light beams up to the speed of light~\citep{GLa15}.
Natural processes, such as stellar explosions~\citep{Loeb20,LL20} or
gravitational slingshot near black hole pairs~\citep{GL15,LG16}, could
launch objects to similar speeds. However, it would be difficult for
relativistic payloads to slow down below the escape speed of Earth,
$10^{-4.5}c$, without having around the same facilities that generated
their high initial speeds.

A better suited propulsion technique that was used in all space
missions from Earth is chemical rockets. Since rockets carry their
fuel, they can navigate to a desired planet and slow down near it.

For a rocket of total mass, $m$, and exhaust speed of the ablated gas
relative to the rocket, $v_{\rm exh}$, momentum conservation implies:
$m\dot{v}=-\dot{m}v_{\rm exh}$, where an overdot denotes a time
derivative. The Tsiolkovsky solution to the rocket
equation~\citep{Tsiol}, $\left({m_{\rm initial}/ m_{\rm
    final}}\right)=\exp\left\{\left({v_{\rm final}-v_{\rm
    initial}}\right)/ v_{\rm exh}\right\}$, implies that for
reasonable fuel-to-payload mass ratio, the final speed $v_{\rm final}$
will only be an order of magnitude larger than the exhaust speed. For
typical chemical propellants with $v_{\rm exh}$ of order a few
km~s$^{-1}$, this tyranny of the rocket equation explains why all
human-made spacecraft reached a speed limit of tens of km~s$^{-1}$ or
$\sim 10^{-4}$c.  Interestingly, this speed is comparable to the
escape speed from the Earth's orbit around the Sun, $v_{\rm esc} \sim
42~{\rm km~s^{-1}}$, making it possible for humanity to launch probes
to interstellar space by taking advantage of the motion of the Earth
around the Sun at $v_{\rm initial}\sim 30~{\rm km~s^{-1}}$. Chemical
propulsion may not be sufficient for probes to escape from the
habitable zone around dwarf stars, like the nearest star Proxima
Centuari~\citep{Loeb18,LL18}.

In summary, chemical propulsion allows escape from the habitable zone
of Sun-like stars and enables slowing down near a destination. The
Ukranian report suggests objects with comparable speeds of up to
15~km~s$^{-1}$.

Devices which need to refuel would favor a habitable planet where
liquid water or combustable organic fuel are available. Planets can be
identified from a distance as they transit their star or through
direct imaging~\citep{Winn23}. Once an Earth-like planet is targeted,
an interstellar device can plunge into its atmosphere. In principle, a
multitude of tiny devices can be released from a mothership that
passes near Earth.

At $v_{\rm final}\sim 10^{-4}c$, a probe would cross twice the
distance of the Sun from the Milky-Way center within a time of $\sim
0.5$ Gyr. The fraction of all Sun-like stars that host Earth-like
planets in their habitable zone is in the range $\sim
3$--$100\%$~\citep{Zink19,Hsu20,Bryson21}.  This implies that
self-replicating probes could reach $\sim 10^{10}$ habitable planets
around Sun-like stars in less than a billion years.

Since most stars formed more than a billion years before the
Sun~\citep{Madau14}, it is possible that other technological
civilizations predated ours by the amount of time needed for their
devices to reach Earth. Here, I point out that any supersonic motion
of such devices through the Earth's atmosphere would inevitably be
accompanied by optical emission.

\section{Unavoidable Optical Emission}

An object with a frontal cross-sectional area $A$, moving at a
supersonic speed, $v$, must create a bow shock in the Earth's
atmosphere and dissipate a mechanical power,
\begin{equation}
P\approx {1\over 2} A\rho_a v^3 = 1.5 {\rm TW} (A/10~{\rm m^2})
(\rho_a/0.3~{\rm kg~m^{-3}}) (v/10~{\rm km~s^{-1}})^3 ,
\label{eq:one}
\end{equation}
where $\rho_a$ is the ambient air density which depends on elevation,
normalized here by its value at an elevation of 10~km.

Data on meteors shows that the fraction of the kinetic power which is
radiated away in the optical band is $\sim 10\%$ (see equation (1) and
figure 2 in \citet{Brown02}), implying an optical luminosity,
\begin{equation}
L_{\rm opt}\approx  150 {\rm GW} (A/10 {\rm m^2})
(\rho_a/0.3~{\rm kg~m^{-3}}) (v/10~{\rm km~s^{-1}})^3 .
\label{eq:two}
\end{equation}
For a path length $\ell$, this luminosity will persist over a period
of time, $\sim 1 {\rm s}\times (\ell/10~{\rm km})/(v/10~{\rm km~s^{-1}})$.

\section{Conclusions}

I conclude that the reported speeds and sizes of the ``Phantom''
objects~\citep{Zhil22}, would have generated fireballs of detectable
optical luminosity at their suggested distances, and so these objects
could not have appeared dark.  However, if the Phantom objects are ten
times closer than suggested, then their angular motion on the sky
corresponds to a physical velocity that is ten times smaller, $v\sim
1.5~{\rm km~s^{-1}}$, and their inferred transverse size would be
$\sim 0.3$--1.2 meters, both characteristic of artillery shells. 

The phantom objects were reported as dark. Their 
minimal cross-section for blocking light inevitably implies that they
must interact with air molecules.

Since $L_{\rm opt}\propto A \times v^3$, the fireball luminosity
scales with inferred distance to the 5th power, and is reduced to a
modest level of a few MW.  If the artillery shells have a frontal
diameter of only 10 cm, then $L_{\rm opt}\sim 10~{\rm kW}$, which at a
distance of $\sim 1$ km would appear extremely faint.

The luminous and variable object at an inferred altitude of $\sim
1,170$ km which was detected through two-site observations above
Ukraine~\citep{Zhil22}, is likely a satellite.

\bigskip
\bigskip
\section*{Acknowledgements}

This work was supported in part by the {\it Galileo Project} at Harvard
University and a grant from the Breakthrough Prize Foundation. 

\bigskip
\bigskip
\bigskip

\bibliographystyle{aasjournal}
\bibliography{t}
\label{lastpage}
\end{document}